\documentclass[12pt]{article}
\usepackage{graphicx}
\usepackage{emulateapj}

\def\lapprox{\hbox{\lower .8ex\hbox{$\,\buildrel < \over\sim\,$}}}
\def\gapprox{\hbox{\lower .8ex\hbox{$\,\buildrel > \over\sim\,$}}}

\begin{document}

\title{Tycho's supernova: the view from {\it Gaia}}

 \author{Pilar Ruiz--Lapuente$^{1,2}$, Jonay I. Gonz\'alez 
Hern\'andez$^{3,4}$, Roger Mor$^{2}$,
 Merc\`e Romero--G\'omez $^{2}$, 
 N\'uria Miret--Roig$^{2,5}$, Francesca Figueras$^{2}$, Luigi R.  
Bedin$^{6}$, Ramon Canal$^{2}$,  Javier M\'endez$^{7}$}

\altaffiltext{1}{Instituto de F\'{\i}sica Fundamental, Consejo Superior de 
Investigaciones Cient\'{\i}ficas, c/. Serrano 121, E--28006, Madrid, Spain}
\altaffiltext{2}{Institut de Ci\`encies del Cosmos (UB--IEEC),  c/. Mart\'{\i}
i Franques 1, E--08028 Barcelon, Spain}
\altaffiltext{3}{Instituto de Astrof\'{\i}sica de Canarias, E--38206 La Laguna,
Tenerife, Spain}
\altaffiltext{4}{Universiad de La Laguna, Departamento de Astrof\'{\i}ica, 
E--38206 La Laguna, Tenerife, Spain}
\altaffiltext{5}{Laboratoire d'Astrophysique de Bordeaux, Universit\'e de 
Bordeaux, CNRS, B18N, All\'ee Geoffroy Saint--Hilaire, 33615 Pessac, France}
\altaffiltext{6}{INAF, Osservatorio Astronomico di Padova, Via dell' 
Osservatorio 3, I--35122 Padova, Italy}
\altaffiltext{7}{Isaac Newton Group of Telescopes, P.O. Box 321, E--38700 
Santa Cruz de La Palma, Spain}

\begin{abstract}

SN 1572 (Tycho Brahe's supernova) clearly belongs to the Ia (thermonuclear) 
type. It was produced by the explosion of a white dwarf in a binary 
system. Its remnant has been the first of this type to be explored in search 
of a possible surviving companion, the mass donor that brought the white dwarf
to the point of explosion. A high peculiar motion with respect to the stars
at the same location in the Galaxy, mainly due to the orbital velocity at the 
time of the explosion, is a basic criterion for the detection of such 
companions. Radial velocities from the spectra of the stars close to the 
geometrical center of Tycho's supernova remnant, plus proper motions of 
the same stars, obtained by astrometry with the {\it Hubble Space Telescope}, 
have been used so far. In addition, a detailed chemical analysis of the 
atmospheres of a sample of candidate stars had been made. However, the 
distances to the stars remained  uncertain. Now, the Second  {\it Gaia}
Data Release (DR2) 
provides unprecedented accurate distances and new  proper motions for the stars 
can be compared with those obtined 
from the {\it HST}.  We consider the Galactic 
orbits that the candidate stars to SN companion  would have in the future.
We do this to explore any kinematic peculiarity. We also locate a representative
sample of candidate stars 
 in the Toomre diagram. Using the 
new data, we reevaluate here the status of the candidates suggested thus far, 
as well as the larger sample of the stars seen in the central region of the 
remnant. 

\end{abstract} 


 \section{Introduction}

\noindent
Type Ia supernovae (SNe Ia) are the calibrated 
standard candles used in the discovery of the accelerated expansion
of the Universe (Riess et al. 1998; Perlmutter et al. 1999) and they
remain a powerful tool in exploring the nature of dark energy. Although
a lot of progress has been made in disentangling the nature of the
explosions, there are still many points to be addressed concerning 
the progenitors (see reviews by Wang \& Han 2012; Maoz et al. 2014, and
Ruiz--Lapuente 2014, for instance). They appear to be thermonucler
explosions of white dwarfs (WDs) made of C+O, and accretion of material 
by the WD from a companion in a close binary system should be the basic 
mechanism to induce the explosion, but here the consensus stops. 
The companion could either be a still thermonuclearly active star 
in any stage of its evolution (the single--degenerate, SD channel) or 
another WD (the double--degenerate, DD channel. The explosion could also
result from the merging of a WD with the electron--degenerate core of 
an asymptotic giant branch (AGB) star. The mode of the accretion could range 
from steady accretion to violent merger, and the explosion either arise from 
central ignition of C, when the WD grows close to the Chandrasekhar mass, or 
be induced by detonation of a He layer near 
the surface, the mass of the WD being smaller in this case. Observed 
different type Ia SNe may have different origins. 

\noindent  
No binary system has ever been discovered in which a SN Ia has 
later taken place, but some binary systems are however considered
to be excellent candidates for SN Ia progenitors, such as U Sco, 
which contains a WD already close to the Chandrasekhar mass. A 
general prediction for the SD channel is that the companion star 
of the WD should survive the explosion and present revealing 
characteristics.
 
\noindent
There are remnants (SNRs) of the explosions of SNe Ia, close and recent 
enough that their exploration can either detect the presence of a surviving 
companion or confirm its absence (Ruiz--Lapuente 1997). 
This has been done for several SNRs of the Ia type, in our own Galaxy and in
the LMC (Ruiz--Lapuente et al. 2004; Gonz\'alez Hern\'andez et al. 2009; 
Kerzendorf et al. 2009; Schaefer \& Pagnotta 2012; Edwards et al. 2012; 
Gonz\'alez Hern\'andez et al. 2012; Kerzendorf et al. 2012, 2013, 2014, 
2018a,b; Bedin et al. 2014; Pagnotta \& Schaefer 2015; Ruiz--Lapuente et al. 
2018). 

\noindent
The remnant of SN 1572 (Tycho Brahe's SN) was the first to be explored 
(Ruiz--Lapuente et al. 2004, RL04 hereafter), and the findings there have 
later been the subject of several studies (Gonz\'alez Hern\'andez et al. 2009,
GH09 henceforth; Kerzendorf et al. 2009; Kerzendorf et al. 2013, 
hereafter K13; Bedin et al. 
2014, B14 hereafter).

\noindent
Now the {\it Gaia} Data Release 2 is providing an unprecedented view
of the kinematics of the Galactic disk (Brown et al. 2018). It not 
only gives the 3D location of a very large sample of stars in the
Galaxy, but also full velocity information (proper motion and 
radial velocity) for 7.2 million stars brighter than 
$G_{\rm RVS}$ = 12 mag, and transverse velocity for an unprecedently
large number of stars. 
{\it Gaia} DR2 provides astrometric parameters (positions, parallaxes and
proper motions) for 1.3 billion sources. The median uncertainty for the
sources brighter than $G$ = 14 mag is 0.03 mas for the parallax and 0.07 
mas yr$^{-1}$ for the proper motions. The reference frame is aligned 
with the International Celestial Reference System (ICRS) and non--rotating 
with respect to the quasars to within 0.1 mas yr$^{-1}$. The systematics 
are below 0.1 mas and the parallax zeropoint uncertainty is small, about
0.03 mas (Brown et al. 2018). 

\noindent
Previously, the distances to the stars could only be estimated from 
comparison of the  absolute magnitudes deduced from their spectral
types and luminosity classes with their photometry, assuming  
some interstellar extinction in the direction of the SNR. That
left considerable uncertainty in many cases  (see RL04 and B14). 
It is here where the {\it Gaia} DR2 is most useful. 

\noindent
The situation was better concerning proper motions, where 
{\it HST} astrometry, based on images taken at different epochs,
had allowed high precision (see B14). {\it HST} proper motions are 
always relative to a local frame, whereas {\it Gaia} DR2
proper motions are absolute, referred to the ICRS. Moreover, {\it Gaia} DR2 allows to
 calculate 
the Galactic orbits of the stars.
In addition, without a precise knowledge
of the distances, the conversion of proper motions into
tangential velocities remained uncertain and so was the reconstruction
of the total velocities.

\noindent
The paper is organized as follows. First we describe the characteristcs
of Tycho's SNR. In Section 3, we examine the distances 
given by the parallaxes from {\it Gaia},
for the surveyed stars, 
and we compare them with previous estimates. In Section 4, the proper motions
from {\it Gaia} are compared with the {\it HST} ones. 
Section 5 discusses the position in the Toomre diagram of possible companion
stars to SN 1572, as compared with a 
large sample. In Section 6, we calculate the Galactic orbits of 4 
representative stars and we discuss their characteristics. In Section 7 our 
whole sample is discussed. Section 8 compares the observations with the 
predictions of models of the evolution of SN Ia companions. Finally,
Section 9 gives a summary and the conclusions.   

\section{Tycho SN remnant}

Tycho's SNR lies close to the Galactic plane ($b$ = 1.4 degrees, 
which means 59--78 pc above the Galactic plane). The remnant
 has angular radius 
of 4 arcmin. In RL04 
a search was performed 
covering the innermost 0.65 arcmin radius centered on the Chandra 
X--ray observatory center of the SN, up to an apparent visual magnitude of
22. Presently we will discuss more stars, roughly doubling the radius of the searched 
area (see Figure 1). The coordinates of the Chandra geometrical center of the remnant are:
RA = 00h 25m 19.9s, DEC = 64o 08' 18.2'' (J2000). 
This is the preferred center, which
pratically coincides with that of ROSAT (Hughes 2000), 
that differs by only 6.5 arcsec. 
The centroid in radio, from VLA (Reynoso et al. 1997), is also nearby. 
The stars 
closest to the center are A, B, C, D, E, F and G. They are the preferred
candidates because of that.

\noindent
The distance to SN 1572 has been subject to study using different 
methods. The estimated value is converging into a value in the middle
of the range from 2 to 4 kpc. Chevalier, Kirshner \& Raymond (1980) 
using the the expansion 
of the filaments in the remnant and the shock velocity obtained a distance
of 2.3$\pm$0.5 kpc. A similar distance was obtained by Albinson et al. 
(1986) through the observation of neutral hydrogen towards the supernova.
They place the distance in the range of 1.7--3.7 kpc. Just one year later, 
Kirshner, Winkler \& Chevalier (1987) revisited the distance through the 
expansion of the filaments of the remnant and found it to be between 2.0 and 
2.8 kpc. 

\bigskip

\noindent
In 2004, Ruiz--Lapuente (2004) attempted a different approach. By assembling
the records of the historical observations of this supernova in 1572--1574
and evaluating the uncertanties, it was possible to reconstruct the light 
curve of the SNIa and the colour. After applying the stretch factor 
fitting of light curves of SNe, it was possible to classify this SN within  
the family of SNe Ia. The derived absolute magnitude  was found 
to be consistent with a distance of 2.8 $\pm$ 0.4 kpc for the scale 
of $H_{0}$ $\sim$ 65 km s$^{-1}$ Mpc$^{-1}$.  In this determination, 
 the extinction towards the supernova was derived from  the 
reddening as shown in the color curve of the SN.
 Given that present  estimates of H$_{0}$ are 67 
km s$^{-1}$ Mpc$^{-1}$, this impacts into a somehow smaller value around 2.7 kpc.

\bigskip

\noindent
With the acknowledgement of  those uncertainties, we take a range of 
possible distance, in this paper, between 1.7 and 3.7 kpc (2.7$\pm$ 1 kpc)
and we study all the stars within this distance range as derived by 
{\it Gaia} as potential candidates. We discuss the distance to the stars in the
next section and we come back to it when talking about candidate stars.
We now see a difference in the
distance towards some stars, and we compare with that  published before.

\bigskip

\begin{figure*}
\centering
\includegraphics[width=0.7\columnwidth]{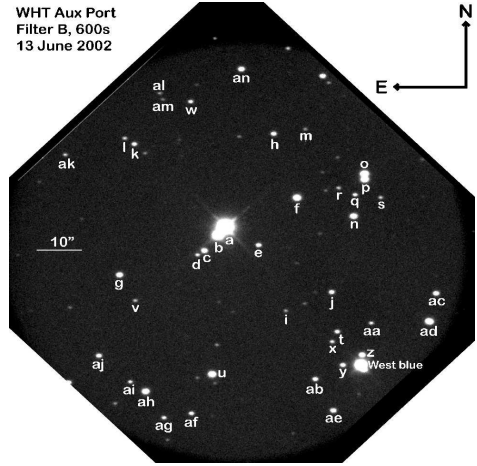}
\caption{$B$--band image, taken with the 4.2m William Herschel Telescope, 
showing all the stars referred to in this paper.}
\label{Figure 1}
\end{figure*}

\section{Parallaxes and {\it Gaia} distances}

\begin{table*}
\scriptsize
\begin{center}
\caption{\scriptsize {\it Gaia} IDs, parallaxes, proper motions and $G$ 
magnitudes of the sample of stars in Figure 1, from the {\it Gaia} DR2}
\begin{tabular}{lccccc}
\\
\hline
\hline
Star &  {\it Gaia} ID & $\varpi$ & $\mu_{\alpha}$ cos $\delta$ & $\mu_{\delta}$ &  $G$  \\ 
     &                &   [mas]  &          [mas/yr]          &   [mas/yr]    & [mag] \\
 (1) &     (2)   &           (3) &            (4)             &     (5)       &  (6)  \\ 
\hline
 A  & 431160565571641856 &  1.031$\pm$0.052 & -5.321$\pm$0.076 & -3.517$\pm$0.065 & 12.404$\pm$0.001 \\
 B  & 431160569875463936 &  0.491$\pm$0.051 & -4.505$\pm$0.063 & -0.507$\pm$0.049 & 15.113$\pm$0.001 \\
 C1 & 431160359417132800 &  5.310$\pm$0.483 & -2.415$\pm$0.735 & -0.206$\pm$0.576 & 18.027$\pm$0.006 \\
 D  & 431160363709280768 &  1.623$\pm$0.318 & -4.566$\pm$0.636 & -2.248$\pm$0.376 & 19.371$\pm$0.003 \\
 E  & 431160565573859584 &  0.138$\pm$0.220 &  0.232$\pm$0.377 & -0.699$\pm$0.265 & 18.970$\pm$0.002 \\
 F  & 431160569875460096 &  0.466$\pm$0.079 & -5.739$\pm$0.130 & -0.292$\pm$0.097 & 17.036$\pm$0.001 \\
 G  & 431160359413315328 &  0.512$\pm$0.021 & -4.417$\pm$0.191 & -4.064$\pm$0.143 & 17.988$\pm$0.001 \\
 H  & 431160599931508480 &  0.620$\pm$0.203 & -4.839$\pm$0.341 & -0.577$\pm$0.248 & 18.895$\pm$0.002 \\
 I  & 431160569867713152 & -0.014$\pm$0.566 & -1.479$\pm$0.970 & -0.855$\pm$0.761 & 20.351$\pm$0.006 \\
 J  & 431160565571749760 &  0.134$\pm$0.240 & -3.900$\pm$0.373 & -1.054$\pm$0.292 & 18.965$\pm$0.002 \\
 K  & 431160393780294144 & -0.266$\pm$0.290 & -1.735$\pm$0.601 & -0.815$\pm$0.350 & 19.313$\pm$0.003 \\
 L  & 431160398076768896 &  0.689$\pm$0.457 & -2.471$\pm$0.876 &  0.514$\pm$0.578 & 20.072$\pm$0.005 \\
 M  & 431160604230502400 & -2.282$\pm$1.316 &  3.472$\pm$1.943 & -1.624$\pm$2.070 & 20.900$\pm$0.011 \\
 N  & 431160565571767552 &  0.246$\pm$0.096 &  0.092$\pm$0.148 &  0.134$\pm$0.121 & 17.612$\pm$0.001 \\ 
 O  & 431160569875457792 &  1.169$\pm$0.063 &  2.607$\pm$0.098 &  2.108$\pm$0.076 & 16.542$\pm$0.001 \\
 P  & 431160565571767424 &  0.168$\pm$0.092 & -0.889$\pm$0.139 & -0.389$\pm$0.106 & 16.998$\pm$0.001 \\
 Q  & 431160565575562240 &  0.663$\pm$0.334 &  0.438$\pm$0.643 &  1.409$\pm$0.404 & 19.496$\pm$0.004 \\
 S  & 431160565573859840 &  1.235$\pm$0.417 &  2.091$\pm$0.771 & -0.437$\pm$0.491 & 19.568$\pm$0.003 \\
 T  & 431159088102994432 &  0.565$\pm$0.289 & -5.177$\pm$0.563 &  0.004$\pm$0.353 & 19.320$\pm$0.003 \\
 U  & 431159092406721280 &  0.504$\pm$0.070 & -1.877$\pm$0.113 & -5.096$\pm$0.083 & 17.064$\pm$0.001 \\
 V  & 431160359413311616 &  0.059$\pm$1.023 & -2.201$\pm$1.184 &  1.645$\pm$1.279 & 20.235$\pm$0.007 \\
 W  & 431160393773079808 &  0.193$\pm$0.283 & -2.760$\pm$0.600 &  0.163$\pm$0.343 & 19.312$\pm$0.003 \\
 X  & 431159092398964992 &  0.192$\pm$0.427 & -1.187$\pm$0.812 & -0.836$\pm$0.511 & 19.812$\pm$0.004 \\
 Y  & 431159092406717568 &  0.631$\pm$0.223 &  0.144$\pm$0.347 & -2.261$\pm$0.290 & 18.923$\pm$0.002 \\
 Z  & 431159092398966400 &  0.176$\pm$0.146 & -1.498$\pm$0.233 & -0.294$\pm$0.193 & 18.082$\pm$0.002 \\
 AA  & 431159088102995968 &  0.957$\pm$0.467 & -2.277$\pm$0.977 & -1.184$\pm$0.595 & 19.973$\pm$0.005 \\
 AB  & 431159088102989056 & -0.090$\pm$0.267 & -2.011$\pm$0.445 & -1.600$\pm$0.316 & 19.046$\pm$0.002 \\
 AC  & 431159088103003520 &  0.490$\pm$0.160 & -2.376$\pm$0.249 & -1.445$\pm$0.195 & 18.399$\pm$0.001 \\
 AE  & 431159088102986880 &  0.279$\pm$0.173 & -0.907$\pm$0.268 & -0.241$\pm$0.227 & 18.559$\pm$0.002 \\
 AF  & 431158881944551424 &  1.323$\pm$0.324 & -2.381$\pm$0.644 &  0.489$\pm$0.400 & 19.399$\pm$0.003 \\
 AG  & 431158881944550272 &  0.703$\pm$0.382 & -2.412$\pm$0.779 &  0.626$\pm$0.453 & 19.768$\pm$0.004 \\
 AH  & 431158881944553216 &  0.206$\pm$0.087 & -0.704$\pm$0.139 & -0.579$\pm$0.107 & 17.486$\pm$0.001 \\
 AI1/HP1 & 431160359417132928 &  2.831$\pm$0.273 & 71.558$\pm$0.530 & -3.030$\pm$0.322 & 19.159$\pm$0.003 \\
 AJ  & 431160359413306368 &  0.187$\pm$0.206 & -1.102$\pm$0.333 &  0.222$\pm$0.249 & 18.883$\pm$0.002 \\
 AK  & 431160393773068032 & -0.476$\pm$0.447 & -1.306$\pm$0.938 & -0.142$\pm$0.518 & 20.008$\pm$0.005 \\
 AL  & 431160398072078592 &  0.383$\pm$0.618 & -2.827$\pm$1.199 &  0.727$\pm$0.789 & 20.461$\pm$0.006 \\
 AM  & 431160398073281792 &  0.752$\pm$0.825 &  0.303$\pm$1.636 & -2.940$\pm$1.114 & 20.605$\pm$0.008 \\
 AN  & 431160599931516288 &  0.605$\pm$0.138 & -4.560$\pm$0.221 & -1.330$\pm$0.168 & 18.284$\pm$0.001 \\
\hline
\end{tabular}
\end{center}
\end{table*}

\begin{table*}
\scriptsize
\begin{center}
\caption{ {\it BVR} photometry, distances and proper motions of 
stars A--W from B14, compared with the distances and proper motions from 
{\it Gaia} DR2 (here the {\it Gaia} proper motions have been transformed to 
the system used in B14; see text and, for the {\it Gaia} original values, 
see Table 1). There are stars that have no upper limit for the {\it Gaia} distances. 
This corresponds to negative parallaxes.  
 They have been marked 
with xxx. There are other stars (I, K, and AB in Table 3) with negative central value of 
the parallax, but adding the errors give postive limits. This corresponds
to a lower limit for the distance, therefore they  
show a $\geq$ sign.}
\begin{tabular}{cccccccccc}
\\
\hline
\hline
Star & {\it B}              &  {\it V}             &  {\it R}               & {\it d} (B14)       & 
{\it d}               & $\mu_{\alpha}$ cos $\delta$ & $\mu_{\alpha}$ cos $\delta$ (B14)& $\mu_{\delta}$&$\mu_{\delta}$ (B14)\\
     & [mag]          & [mag]         & [mag]           & [kpc]             &
[kpc]                 & [mas/yr]      & [mas/yr]        & [mas/yr]          & [mas/yr]\\
(1)    & (2)            & (3)           & (4)             & (5)               &
(6)                   & (7)           & (8)             & (9)               & (10)    \\      
\hline
A    & 14.82$\pm$0.03 & 13.29$\pm$0.03 & 12.24$\pm$0.03 & 1.1$\pm$0.3 &
0.97$^{+0.05}_{-0.04}$  & -3.63$\pm$0.08  & ---  & -3.06$\pm$0.07       & ---    \\
B    & 16.35$\pm$0.03 & 15.41$\pm$0.03 & ---            & 2.6$\pm$0.5&  
2.03$^{+0.19}_{-0.15}$ & -2.90$\pm$0.06  & -1.67$\pm$0.06 &-0.09$\pm$0.05 & 0.59$\pm$0.08\\
C1   & 21.06$\pm$0.12 & 19.06$\pm$0.05 & 17.77$\pm$0.03& 0.75$\pm$0.5 &
0.18$^{+0.03}_{-0.01}$ & -0.82$\pm$0.73  & -1.98$\pm$0.07 & 0.39$\pm$0.58 &-1.09$\pm$0.06\\ 
C2   & 22.91$\pm$0.20 & 20.53$\pm$015  & ---   & $\sim$40     &
---                  & ---             & -1.75$\pm$0.07   & ---         &-1.07$\pm$0.07\\
C3   & ---           & ---             & ---   & ---  &
---  & ---           & 0.08$\pm$0.11   & ---   & -0.14$\pm$0.10\\      
D    & 22.97$\pm$0.28 & 20.70$\pm$0.10 & 19.38$\pm$0.06& 0.8$\pm$0.2  &
0.62$^{+0.15}_{-0.11}$& -2.97$\pm$0.64   & -2.03$\pm$0.09 &-1.65$\pm$0.38 &-1.28$\pm$0.07\\
E    & 21.24$\pm$0.13 & 19.79$\pm$0.07 & 18.84$\pm$0.05& $>$20        &
7.22$^{+xxx}_{-4.43}$& 1.83$\pm$0.38    &  1.74$\pm$0.05 &-0.10$\pm$0.26 & 0.28$\pm$0.05\\ 
F    & 19.02$\pm$0.05 & 17.73$\pm$0.03 & 16.94$\pm$0.03 & 1.5$\pm$0.5 &
2.15$^{+0.44}_{-0.32}$& -4.14$\pm$0.13   & -3.31$\pm$0.15 & 0.31$\pm$0.10 & 0.25$\pm$0.07\\
G    & 20.09$\pm$0.08 & 18.71$\pm$0.04 & 17.83$\pm$0.03 & 2.5-5.0     &
1.95$^{+0.60}_{-0.35}$ & -2.82$\pm$0.19  & -2.63$\pm$0.06 &-3.46$\pm$0.14 &-3.98$\pm$0.04\\
H    & 21.39$\pm$0.14 & 19.80$\pm$0.07 & 18.78$\pm$0.05 & $\simeq$1.8/$\sim$24 &
 1.61$^{+0.79}_{-0.40}$  & -3.24$\pm$0.34 & -3.13$\pm$0.07 &-0.02$\pm$0.25 &-0.84$\pm$0.03\\
I    & ---            & 21.75$\pm$0.16 & 20.36$\pm$0.09 & $\simeq$4  &
$\geq$ 1.81                   &  0.12$\pm$0.97 &  0.69$\pm$0.06 &-0.25$\pm$0.76 &-0.20$\pm$0.06\\ 
J    & 21.15$\pm$0.12 & 19.74$\pm$0.07 & 18.84$\pm$0.05 & $\simeq$9  &
7.46$^{+xxx}_{-4.77}$  & -2.30$\pm$0.37 & -2.35$\pm$0.06 &-0.45$\pm$0.29 &-0.28$\pm$0.03\\ 
K    & 21.64$\pm$0.15 & 20.11$\pm$0.08 & 19.15$\pm$0.05 & $\simeq$2.4/$\sim$27 &
$\geq$ 41.67           & -0.14$\pm$0.60 &  0.24$\pm$0.12 &-0.21$\pm$0.35 & 0.03$\pm$0.07\\  
L    & 22.77$\pm$0.26 & 21.08$\pm$0.12 & 20.00$\pm$0.07 & $\simeq$4  &
1.45$^{+2.87}_{-0.58}$  & -0.87$\pm$0.88 &  0.36$\pm$0.12 &1.11$\pm$0.58  &-0.08$\pm$0.04\\
M    & 23.49$\pm$0.36 & 21.82$\pm$0.16 & 20.72$\pm$0.10 & $\simeq$4  &   
---                   & 5.07$\pm$1.94  & -0.61$\pm$0.12 &-1.02$\pm$2.07 & 0.44$\pm$0.08\\
N    & 19.59$\pm$0.06 & 18.29$\pm$0.04 & 17.47$\pm$0.03 & $\simeq$1.5-2 &
4.06$^{+2.57}_{-1.14}$  &  1.69$\pm$0.15  &  2.64$\pm$0.13 &0.74$\pm$0.12  & 0.96$\pm$0.04\\
O    & 18.62$\pm$0.04 & 17.23$\pm$0.03 & 16.37$\pm$0.03 & $<$1 &
0.85$^{+0.54}_{-0.22}$  & 4.21$\pm$0.10  &  5.13$\pm$0.20 &2.71$\pm$0.08  & 2.85$\pm$0.14\\
P1   & ---            & 17.61$\pm$0.03 & 16.78$\pm$0.03 & $\simeq$1 &
5.96$^{+7.23}_{-2.11}$  & ---            &  1.39$\pm$0.36 & ---           & 0.20$\pm$0.09\\
P2    & ---           & ---            & ---            & ---       &
---                   & ---            & -0.27$\pm$0.20 & ---           &-1.64$\pm$0.21\\
Q    & 22.35$\pm$0.21 & 20.59$\pm$0.09 & 19.41$\pm$0.06 & $\simeq$2 &
1.51$^{+1.53}_{-0.51}$  & 2.04$\pm$0.64  &  1.34$\pm$0.09 &2.71$\pm$0.40  & 2.38$\pm$0.04\\
R    & 22.91$\pm$0.28 & 21.38$\pm$0.13 & 20.26$\pm$0.08 & 3.3$\pm$0.2 &
---                   & ---            & -0.18$\pm$0.10 & ---           & 0.25$\pm$0.05\\
S    & ---            & 21.30$\pm$0.13 & 19.74$\pm$0.07 & 1.3$\pm$0.1 &         
0.81$^{+0.41}_{-0.20}$  & 3.69$\pm$0.77  &  3.68$\pm$0.09 & 0.16$\pm$0.49 & 0.93$\pm$0.05\\
T    & 21.82$\pm$0.17 & 20.23$\pm$0.08 & 19.20$\pm$0.05 & $\simeq$2/$\sim$30 &
1.77$^{+1.86}_{-0.60}$  &-3.58$\pm$0.56  & -2.96$\pm$0.04 & 0.61$\pm$0.35 &-0.53$\pm$0.05\\
U    & 19.03$\pm$0.05 & 17.73$\pm$0.03 & 16.95$\pm$0.03 & $\simeq$1  &
1.98$^{+0.32}_{-0.24}$  & -0.28$\pm$0.11 &  0.39$\pm$0.10 &-4.49$\pm$0.08 &-4.31$\pm$0.07\\
V    & 23.32$\pm$0.33 & 21.41$\pm$0.13 & 20.20$\pm$0.08 & $\simeq$3  &
16.81$^{+xxx}_{-15.89}$& -1.16$\pm$1.18 & -0.67$\pm$0.08 &2.25$\pm$1.28  & 0.49$\pm$0.08\\
W    & 22.13$\pm$0.19 & 20.44$\pm$0.09 & 19.27$\pm$0.05 & $\simeq$2  &
5.17$^{+xxx}_{-3.07}$  & -1.16$\pm$0.60 & -0.31$\pm$0.09 &0.76$\pm$0.34  & 0.09$\pm$0.04\\
\hline 

\end{tabular}
\end{center}
\end{table*}

\begin{table*}
\scriptsize
\begin{center}
\caption{ $G$ magnitudes, distances and proper motions of stars X--AN from B14, 
with comparison of the proper motions from B14 and from {\it Gaia} DR2 (here 
the {\it Gaia} proper motions have been transformed to the system used in B14; 
for the {\it Gaia} original values, see Table 1). These stars have not 
been assigned a distance in B14 nor in any other paper. The sign xxx for the
 upper
limit in distance to some stars correspond to negative upper limit parallax, as in Table 2.}
\begin{tabular}{ccccccc}
\\
\hline
\hline
Star & {\it G} & {\it d} & $\mu_{\alpha}$ cos $\delta$ & $\mu_{\alpha}$ cos $\delta$ (B14) & $\mu_{\delta}$ & $\mu_{\delta}$ (B14)  \\
     & [mag]       & [kpc] & [mas/yr] & [mas/yr] & [mas/yr] & [mas/yr]  \\
(1)  & (2)         & (3)   & (4)      & (5)      & (6)      & (7)       \\      
\hline
X    & 19.81          & 5.20$^{+xxx}_{-3.59}$ &  0.41$\pm$0.81  & 1.32$\pm$0.06 & -0.24$\pm$0.51 & -0.26$\pm$0.06 \\
Y    & 18.92          & 1.58$^{+0.87}_{-0.41}$ &  1.74$\pm$0.35  & 2.52$\pm$0.03 & -1.66$\pm$0.29 & -1.75$\pm$0.07 \\
Z    & 18.08          & 5.68$^{+2.72}_{-2.57}$ &  0.10$\pm$0.23  & 0.78$\pm$0.04 &  0.31$\pm$0.19 &  0.06$\pm$0.07 \\
AA   & 19.97          & 1.04$^{+1.00}_{-0.33}$ & -0.68$\pm$1.00  &-3.19$\pm$0.07 & -0.58$\pm$0.60 & -1.42$\pm$0.04 \\
AB   & 19.05          & $\geq$ 5.65                & -0.41$\pm$0.45  &-0.37$\pm$0.03 & -1.00$\pm$0.32 & -1.01$\pm$0.06 \\
AC   & 18.40          & 2.04$^{+0.99}_{-0.50}$ & -0.78$\pm$0.25  &-1.09$\pm$0.07 & -0.84$\pm$0.20 & -0.87$\pm$0.04 \\
AD1  & 17.22          & 1.04$^{+0.84}_{-0.32}$ &  0.85$\pm$0.59  &-1.24$\pm$0.14   & ---           & -1.58$\pm$0.16 \\
AD2  & ---            & ---                   & ---             &-1.12$\pm$0.04 & ---           & -2.25$\pm$0.07 \\
AE   & 19.05          & 3.58$^{+5.84}_{-1.37}$  & 0.69$\pm$0.27   & 1.07$\pm$0.06 & 0.36$\pm$0.23 & -0.05$\pm$0.06 \\
AF   & 19.40          & 0.76$^{+0.24}_{-0.15}$  &-0.78$\pm$0.64  &-0.38$\pm$0.04 &  1.09$\pm$0.40 &  0.01$\pm$0.07 \\
AG   & 19.77          & 1.42$^{+1.69}_{-0.50}$ & -0.81$\pm$0.78  &-1.11$\pm$0.06 &  1.23$\pm$0.45 &  0.93$\pm$0.08 \\
AH   & 17.49          & 4.85$^{+3.55}_{-1.44}$ &  0.89$\pm$0.14  & 1.26$\pm$0.20 &  0.02$\pm$0.11 & -0.20$\pm$0.26 \\
AI1/HP-1& 19.16       & 0.35$^{+0.04}_{-0.03}$ & 73.16$\pm$0.53  &73.07$\pm$0.09 & -2.43$\pm$0.32 & -2.82$\pm$0.07 \\
AI2  & ---            & ---                  & ---             & 1.76$\pm$0.28 & ---            &  0.16$\pm$0.21 \\
AJ   & 18.88          & 5.35$^{+xxx}_{-2.81}$ &  0.50$\pm$0.33  & 0.18$\pm$0.05 &  0.82$\pm$0.25 &  0.73$\pm$0.07 \\
AK   & 20.01          & ---                  &  0.29$\pm$0.94  &-0.25$\pm$0.09 &  0.46$\pm$0.52 &  0.95$\pm$0.08 \\
AL   & 20.46          & 2.61$^{+xxx}_{-1.61}$ & -1.23$\pm$1.20  &-0.14$\pm$0.09 &  1.33$\pm$0.73 & -0.35$\pm$0.09 \\
AM   & 20.61          & 1.33$^{+xxx}_{-0.70}$ &  1.90$\pm$1.64  &-1.36$\pm$0.10 & -2.34$\pm$1.11 & -0.60$\pm$0.10 \\
AN   & 18.28          & 1.65$^{+0.49}_{-0.30}$ & -2.96$\pm$0.22  &-2.83$\pm$0.11 & -0.73$\pm$0.17 & -0.96$\pm$0.05 \\   
\hline  

\end{tabular}
\end{center}
\end{table*}

\noindent
Distances to the stars targeted as possible surviving companions of 
SN 1572 had first been estimated by RL04 (see their Table 1), for
13 of them.
Those estimates were made by fitting synthetic spectra (under the
assumption of local thermodynamic equilibrium, LTE) to the observed 
ones. The grids of model atmospheres and the atomic data from 
Kurucz (1993), in combination with the Uppsala Synthetic Spectrum 
Package (1975), were used in the  spectrum synthesis. The atmospheric 
parameters effective temperature $T_{\rm eff}$ and surface gravity 
$g$ were thus determined. Intrinsic colours and absolute visual 
magnitudes were then deduced from the relationships between spectral 
type and colour and spectral type and absolute magnitude for the 
different luminosity classes (Schmidt--Kahler 1982). Comparison with 
$BVR$ photometry obtained with the 2.5m Isaac Newton Telescope, in
La Palma, yielded the reddening $E(B - V)$, from which the visual 
extinction $A_{V}$ and the corrected apparent visual magnitude 
$V_{0}$ were calculated. The high--resolution spectra had been 
obtained with the UES and ISIS spectrographs, in the 4.2m William 
Herschel Telescope, in La Palma. Low--resolution spectra came, in 
addition, from the LRIS imaging spectrograph in the 10m Keck 
Telescopes, in Hawaii. They were compared, after dereddening, with 
template spectra from Lejeune et al. (1997), and that supplemented 
the information obtained from the high--resolution spectra.

\noindent
The detailed characterization of Tycho G (singled out as a likely
SN companion in RL04) was done by GH09 using a high--resolution 
HIRES spectrum obtained at the Keck I telescope. The stellar 
parameters, effective temperature and gravity, were derived using 
the excitation and ionization equilibria of Fe together with the fit 
of the wings of the H$\alpha$ line compared to different synthetic 
spectra computed. The result pointed to a G2 IV star with metallicity
slightly below solar. The individual magnitudes in the 
different filters  were  used to
estimate a range of possible distances of star G. In addition, 
 low--resolution LIRIS spectra of the stars E, F, G, and D were obtained,
which confirmed the spectral types of these stars. The best fit 
for Tycho G gave T$_{\rm eff}$ = 5900 $\pm$ 150\ K, log $g$ = 3.85 
$\pm$ 0.35\ dex, and [Fe/H] = -0.05 $\pm$ 0.09 (see GH09). This result 
is consistent with that obtained by K13.

\noindent   
 K13 recalculated spectrophotometric distances to 5 of the stars 
(A, B, C, E and G). Many of them had  large error bars and are compatible
with the distance estimate in B14 and the distance values 
implied by {\it Gaia} parallaxes, except for
the value given for their star C (star C is in fact three  stars. Star C1 has 
a measured {\it Gaia} parallax corresponding to a distance of
 d $=$ 0.18$^{+0.3}_{-0.1}$  kpc. The estimate in B14 is compatible with that 
value, whereas K13 found a too large distance of 5.5$\pm$3.5 kpc.)
 Finally,
in B14 there is a list of distances to 23 stars (A to W) (their Table 3), 
completing the work of RL04. The distances in B14 are in  reasonable 
 agreeement (within 1 $\sigma$) with the {\it Gaia} parallaxes (see Table 2).

\noindent
Now the DR2 from {\it Gaia} has provided us with precise parallaxes 
for almost all the stars included in those  previous studies. The
 corresponding 
distances and their errors are given in columnn 6 of Table 2, for
the stars for which we already had estimates of the distances 
(column 5), and in column 3 of Table 3 for those for which there
were none. {\it Gaia} DR2 distances are estimated as the inverse of the
parallax. 

\begin{figure*}
\centering
\includegraphics[width=0.9\columnwidth]{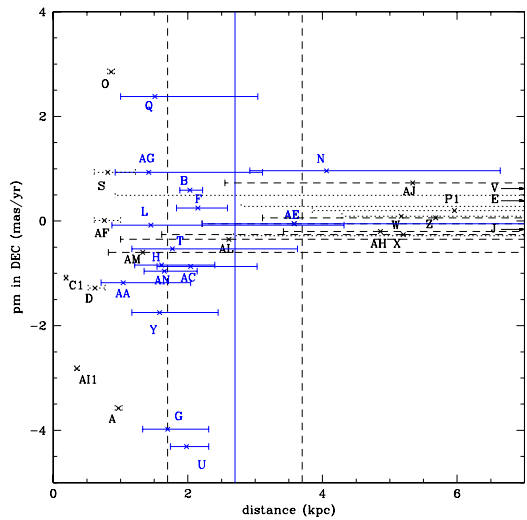}
\caption{Distances and distance ranges inferred from the 
parallaxes in the {\it Gaia} DR2 and their uncertainties, together with their 
proper motions in declination. The dashed vertical lines mark the conservative
limits 
of 2.7 $\pm$ 1 kpc  on the distance to Tycho's SNR. Solid (blue) 
error bars correspond to stars that, within reasonable uncertainties, might be 
inside the SNR, dashed lines to those that, although formally their 1 $\sigma$ 
error bars 
reach the distance of the remnant, they are so large as to make it 
implausible, while dotted lines correspond to the stars 
that are beyond the 
limits for the SNR distance or  have 
a parallax relative error higher than 100$\%$.}
\label{Figure 2}
\end{figure*}

\noindent
In Figure 2 are shown the {\it Gaia} DR2 distances and their 
error bars, and they are compared with the estimated distance to 
Tycho's SNR (blue vertical line) and its error bars (black dashed
vertical lines). Proper motions in declination are on the vertical 
axis. Solid (blue) error bars mark the stars that, 
within reasonable uncertainties, might be inside Tycho's SNR. 
Dashed (black) error bars correspond to stars that, although the error 
bars of their distances reach the range of distances to the 
SNR, the errors are so large as to make their association with
the SNR very unlikely. Finally, dotted (black) lines correspond to the
stars that have incompatible distances with the SNR or parallax errors 
larger than 100$\%$.

\noindent
 There are 15 stars within the range of possible distances 
to the SNR (1.7 $<$ d $<$ 3.7 kpc). These are stars B,  F, 
 G, H,  L, N, Q, T, U, Y, AA, AC, AE, AG and AN.
From this list of stars, stars G and U  have significant
 proper motions in declination. Proper motions
of all the targeted stars will be discussed in the next Section.

\noindent
In Table 1, the {\it  Gaia} DR2 data on parallaxes, proper 
motions, and magnitudes {\it G} are given as in the {\it Gaia} DR2. 
The proper motions  are absolute, referred in the ICRS (as mentioned above).

\noindent
In Table 2, the {\it Gaia} DR2 distances, as we announced,
 are compared with the 
distances deduced in B14 from determination of the 
stellar atmospheric parameters and comparison of the resulting 
luminosities with the available photometry. We see that there
is  reasonable agreement in most cases, with a tendency to place the
stars at longer distances in B14 as compared with {\it Gaia}, 
which can be attributed to an underestimate, in B14, of the 
extinction in the direction of Tycho's SNR. Based only on the
stars with distance errors $\leq$ 0.5 kpc in both sets, 
the underestimate would be by $\Delta A_{V} \simeq 0.5$ mag.
 Exceptions are stars N, P1, U, V and W, although in the two 
latter cases the {\it Gaia} error bars are so large that the 
comparison is not really meaningful.
 
\section{Proper motions  from {\it Gaia} compared with the {\it HST}}

\noindent 
High velocities, mainly due to their orbital motion at the time of explosion, 
must be a salient characteristic of SNe Ia companions. Unless they were mostly
moving along the line of sight when the binary system was disrupted, the 
components of the velocity on the plane of the sky should be observed as high 
proper motions relative to the stars around the location of the SN.
 It might happen that the star were moving at some angle with the line 
of sight and thus the components of the velocity would be distributed accordingly.

\noindent
The
location of the supernova explosion, in the case of SNe whose remnants still exist, is given in a first, 
rough approximation, by the centroid of the remnant. High--precision 
astrometric measurements of the proper motions of the stars within some 
angular distance from the centroid are the tool needed to detect or discard 
the presence of possible companions. Until the advent of {\it Gaia}, this was 
only possible with {\it HST} astrometry. Radial velocities obtained from 
high resolution spectra give the complementary information on the velocity 
along the line of sight.

\noindent
A first set of measurements of the proper motions of the stars around the 
centroid of Tycho's SNR was made in RL04. It included 26 stars, labelled from 
A to W (see their Fig. 1). Images from the WFPC2 aboard the {\it HST}, taken 
two months apart (within Cycle 12), were used. It was  found  that star G 
was at compatible distance to the supernova (the distance estimate
 for Tycho G, (most widely named star G) 
was 3.0 $^{+1}_{-0.5}$ kpc and its motion was mostly
 perpendicular to the Galactic plane, 
with $\mu_{b}$ = 6.11$\pm$1.34 mas/yr ($\mu_{l}$ = -2.6$\pm$1.34 mas/yr only). 
That meant a heliocentric
 tangential velocity $v_{t} \sim$94 km/s, which combined with a high
measured radial velocity of -87.4 $\pm$ 0.5 km/s it gave a total heliocentric 
velocity
$v_{tot}$ $\sim$128 km/s $\pm$ 9 km/s, 
making star G a likely candidate to have been the companion 
star of SN 1572.

\noindent
In B14, proper motions $\mu_{\alpha}$ cos $\delta$ = -2.63$\pm$0.18 mas/yr, and 
$\mu_{\delta}$ = -3.98$\pm$0.10 mas/yr were measured. That, for a distance of
2.83$\pm$0.79 kpc, taken as the SN distance, give a heliocentric tangential velocity
 $v_{t}$ = 64$\pm$11 km/s. Given the heliocentric radial
velocity,  $v_{r}$ = -87.4$\pm$0.5, 
the total velocity is $v_{tot}$ = 108$\pm$9 km/s.

\noindent  
Now, from the {\it Gaia} DR2, $\mu_{\alpha}$cos $\delta$ = -4.417$\pm$0.191 
mas/yr and $\mu_{\delta}$ = -4.064$\pm$0.143 mas/yr. The parallax being 
$\varpi$ = 0.512$\pm$0.021, we have
$v_{\alpha}$ cos $\delta$ = -40.88$\pm$2.44 km/s, 
 $v_{\delta}$ = -37.61$\pm$2.03 km/s, and $v_{t}$ = 55.55$\pm$2.26 km/s
 (heliocentric). 

\smallskip
\noindent
For the corresponding $v_{r}$, it results a $v_{tot}$ = 103.69 $\pm$ 7.52 km/s.
We thus see that, as compared with B14, there is no significant change with the
new results. As in B14, they can be interpreted in the framework of a
binary model similar to that for U Sco. The excess velocity of star G with
respect to the average of the stars at the same location in the Galaxy could come
from the orbital velocity it had when the binary was disrupted by the SN
explosion.

\smallskip
\noindent
Taking as a reference the Besan\c con model of the Galaxy (Robin et al. 2003),
the average heliocentric tangential velocity of disc stars, at the position and distance
of star G, is almost negligible, while the average radial velocity, 
is $\langle v_{r}\rangle \approx -31
\pm 28$ km/s. Then, attributing the excess over average in radial velocity
($\approx$ -56 $\pm$ 28 km/s) to orbital motion and the full tangential
velocity ($\approx$ 55 $\pm$ 2 km/s) to it, we obtain that $v_{orb} \approx$ 78
$\pm$ 20 km/s (the inclination of the plane of the orbit with respect to the
line of sight would thus be $i$ = 44$^{o}$).

\smallskip
\noindent
The evolutionary path giving rise to SN 1572 might have started from a WD
with a mass $\sim$ 0.8 M$_{\odot}$ plus a somewhat evolved companion of
$\sim$ 2.0-2.5 M$_{\odot}$ filling its Roche lobe (RL04),
the system ending up as a WD with the Chandrasekhar mass ($\sim$ 1.4
M$_{\odot}$), plus a companion of $\sim$ 1 M$_{\odot}$. Using Kepler's law,
that $P^{2} =  a^{3}/(M_{1}+M_{2})$ (with $P$ in years, $a$ in astronomical
units, $M_{1}$ and $M_{2}$ in solar masses), and since $P = 2\pi a/v$, we find
a separation $a \approx\ 25 R_{\odot}$, the period being $P \approx 9$ days.
Using Eggleton's (1983) formula

$$ R_{L} = a \left[{0.49\over 0.6 + q^{-2/3} {\rm ln}(1 + q^{1/3})}\right]$$

\noindent
($q$ being the mass ratio $M_{2}/M_{1}$) for the effective Roche lobe radius of
the companion just before the explosion, it would thus have been $\approx$
9 R$_{\odot}$. At present, the radius of the star is only of 1-2 R$_{\odot}$
(GH09), and would have resulted from the
combination of mass stripping and shock heating by the impact of the SN
ejecta, plus subsequent fast cooling of the outer layers up to the present
time.

\smallskip
\noindent
However, an alternative to this hypothesis is discussed later in the paper.

\noindent
In B14, the proper motions of 872 stars were measured from
$HST$ astrometry, using images taken in up to four different epochs 
and spanning a total of 8 yr. Much higher precision than in RL04 was achieved. 
The results for 45 of them (all the stars with names in Figure 1) are given 
in Table 2 and Table 3 of B14. The full version was provided as 
supplementary electronic 
material.

\noindent
When comparing the proper motions given by the {\it Gaia} DR2 with those 
obtained by B14 from the astrometry done with the {\it HST}, one must take 
into account that the former are {\it absolute} measurements, in the ICRS 
system, while the latter are {\it relative} measurements. This means that the 
local frame used for the {\it HST} astrometry should, in general, move with 
respect to the ICRS frame. Such systematic effect is actually seen when we 
make the comparison. In Tables 2 and 3, the {\it Gaia} proper motions have 
been transformed to the {\it HST} frame. 

\noindent
Including only the stars with proper motion errors smaller than 0.25 
mas/yr in B14, we find that, on average, 
$\mu_{\alpha}\
 {\rm cos}\ {\delta}\ (Gaia) = \mu_{\alpha} {\rm cos}\ {\delta}\ (B14) 
- 1.599\pm0.729\ {\rm mas/yr}$
and
$\mu_{\delta}\ (Gaia) = \mu_{\delta}\ (B14) - 0.601\pm0.585\ {\rm mas/yr}$.

\noindent
In Table 2 (columns 7 and 9) and Table 3 (columns 4 and 6), we have transformed 
the {\it Gaia} proper motions to the B14 {\it HST} frame according to these  
relations. For our purposes, the local, relative proper motions are most 
meaningful, since we are interested in the motions of the stars with respect to 
the average motions of those around their positions. After applying these 
zero--point shifts, there still are residual differences between the two proper 
motion sets. On average,
$\Delta\ \mu_{\alpha}\ {\rm cos}\ \delta = -0.017\pm0.788\ {\rm mas/yr}$
and
$\Delta\ \mu_{\delta} =  0.005\pm0.630\ {\rm mas/yr}$. 
The whole set is included here, the dispersion being mainly due to stars
which have substantial errors in their {\it Gaia} proper motions (see 
columns 7 and 9 in Table 2 and columns 4 and 6 in Table 3, as well as 
columns 5 and 6 in Table 4).

\section{Toomre diagram}

\noindent
{\it Gaia} provides a five--parameter astrometric
solution and for some stars line--of--sight velocities 
($\alpha$, $\delta$, $\varpi$, 
$\mu^{*}_{\alpha}$, $\mu_{\delta}$, V$_{r}$), together with their associated 
uncertainties and correlations between the astrometric quantities.
For the 13 stars for which we also know their radial velocities, the 
total space velocities can be derived. It is most useful to see their
components in the Galactic coordinate system: U (positive in the 
direction of the Galactic center), V (positive in the direction of 
Galactic rotation) and W (positive in the direction of the North Galactic 
Pole) in the LSR. In Table 4 we give the U, V and W components of the space velocities, 
as well as the total velocities on the Galactic meridian plane, in the Local 
Standard of Rest, of these 13 stars, based on the {\it Gaia} DR2 parallaxes 
and proper motions and on the radial velocities from B14 (save for star A, 
which has a quite precise radial velocity from {\it Gaia}). For the 
transformation of the  motions from heliocentric to the LSR, we
have adopted, as the peculiar velocity of the Sun with respect to the 
LSR, (U$_{\odot}$, V$_{\odot}$, W$_{\odot}$) = (11.1, 12.24, 7.25) km s$^{-1}$ 
(Sch\"onrich et al. 2010).

\begin{figure*}
\centering
\includegraphics[width=1.0\columnwidth]{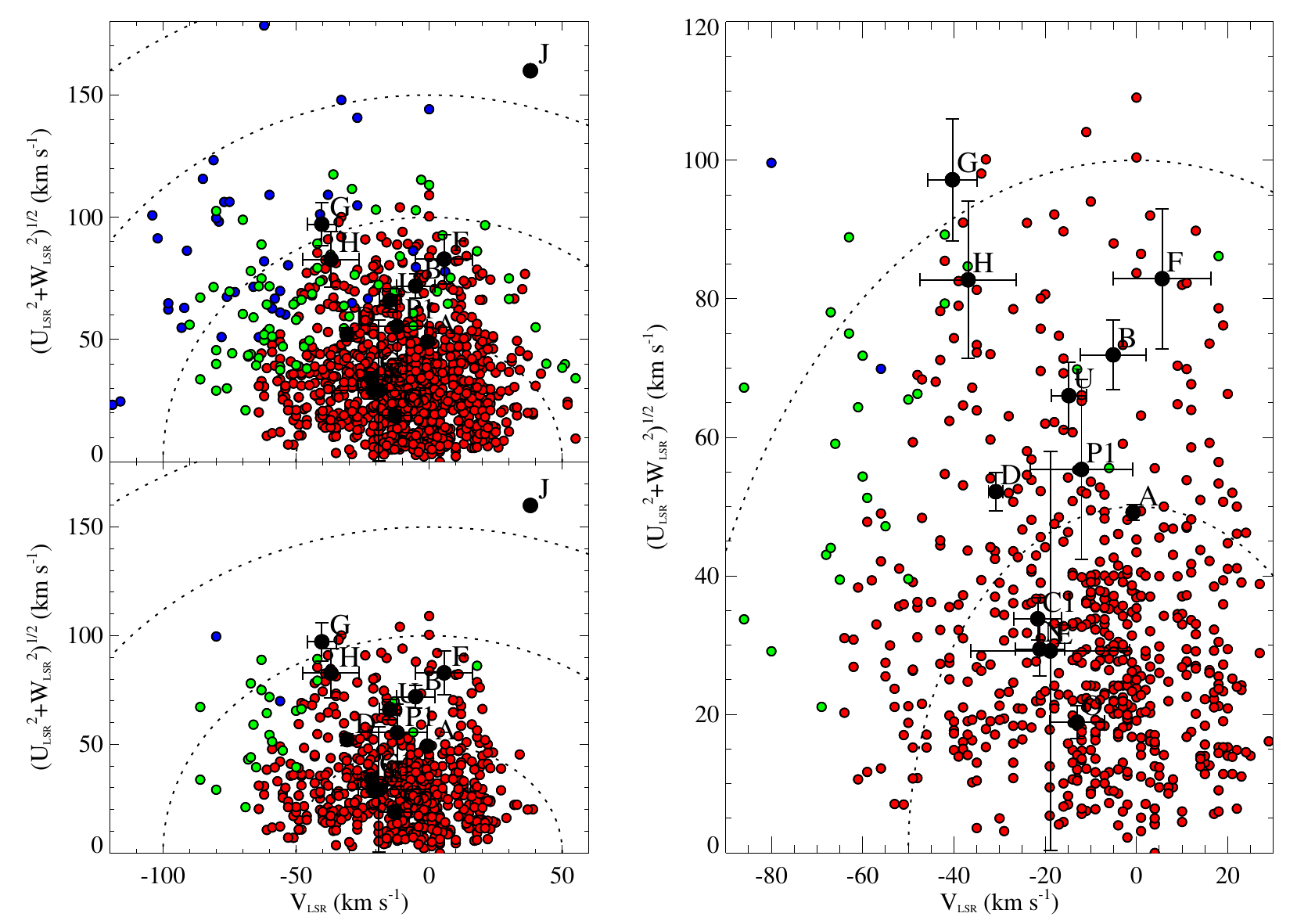}
\caption{
Left upper panel: Toomre diagram for a sample of thin 
disk, thick disk, and transition thin--thick disk stars, covering a wide range 
of metallicities, with our stars (from Table 4)
 superimposed (red dots correspond
to thin disk stars, green to transition, and blue to thick disk 
stars). Left lower panel: same as upper panel, keeping only stars with
metallicities equal to or higher than that of star G. Right panel: detail 
of the lower left panel, leaving out star J. The sample is taken from
Adibekyan et al. (2012) (see text).}

\label{Figure 3}
\end{figure*}

\noindent
 The Toomre diagram shows the distribution of U,V,W 
velocities in the LSR. This diagram 
 combines quadratically U and W versus V and allows to distinguish 
between stars belonging to different Galactic stellar components (thin disk, 
transition thin–-thick, thick disk, and halo). 
 Adibekyan et al. (2012)  sample, with very high quality
spectroscopic data from the HARPS exoplanet search program, is used as a 
reference
in Figure 3.  The data have 
very high precision on radial velocities, stellar parameters and metallicities.
Therefore, the Toomre diagram obtained from this 
sample combines kinematics information such as orbital motion 
together with the information derived from the spectroscopic analysis on 
 element abundances. The sample covers a wide range of metallicities 
[Fe/H] from -1.2 to 0.4.

\noindent
The sample in the upper left panel of Figure 3 
 has no imposed boundaries on
metallicity, while those in the lower left and the right panels 
include only stars with metallicities equal to or higher than that 
of star G minus the 1\ $\sigma$ uncertainty, i.e. for [Fe/H] $>$ 
-0.14. One sees there that (with the exception of star J, 
whose kinematics is very unreliable, with large errors in the 
{\it Gaia} DR2 data), no other star in our sample moves as fast
as star G. 

\smallskip

\noindent
The Gaia DR2 data place star G above the region where most thin disk  
stars are. The kinematics of star G would locate it among the  thick 
disk stars but its metallicity is that of a thin disk star, 
while at its location, only 48 pc above the Galactic plane, the
density of thick disk stars is very low. Using the Adibekyan 
et al. (2012) sample, the probability that star G belonged to the
thick disk, given its metallicity, is only of 2 $\%$.

\smallskip

\noindent
There are some thin disk stars, however, that move fast on the 
Galactic meridian plane, and thus star G might belong to this 
group, although, as we will see, it includes only a small 
fraction of the thin disk stars. 

\noindent
Quantitatively, in the sample from Adibekyan et al. (2012), of 
1111 FGK dwarf stars, there are 601 thin disk stars with 
metallicities [Fe/H] $>$ -0.14. 446 of them (74,2\%) are inside 
the circle {\bf $V^{2} + (U^{2} + W^{2})^{1/2} <$} 50 km/s, and 596
are inside the $<$ 100 km/s circle. Only 5 (0.8\%) have 
velocities higher than 100 km/s. That is, therefore, (0.8\%), the 
probability, from kinematics alone, that star G were just a
fast--moving thin disk star.  

\noindent
In Section {\bf 6} the orbits of the stars are discussed and 
the question of the detailed chemical abundances of star G
will be addressed.

\begin{table*}
\scriptsize
\setlength{\tabcolsep}{4pt}
\begin{center}
\caption{Parallaxes, heliocentric radial velocities, proper motions, Galactic U, V, W velocity components, and 
total velocities on the Galactic meridian plane (referred to the LSR) of the stars with both radial  velocities 
from B14 and parallaxes from {\it Gaia} DR2 (the proper motions, here, are in the {\it Gaia} system).}
\begin{tabular}{llcccccccc}
\\
\hline
\hline
Star & DR2 number & $\varpi$  &  $v_{\rm r}$ &$\mu_{\alpha}\ {\rm cos}\ \delta$&$\mu_ {\delta}$& U & V & W &$(U^{2} + W^{2})^{1/2}$ \\  
 & (4311 ...) &  (mas)&  (km/s)      & (mas/yr)                           & (mas/yr)        & (km/s) & (km/s) & (km/s) & (km/s) \\
(1) & (2) &  (3)  &  (4)         & (5)                                & (6)             & (7)    & (8)    & (9)    & (10) \\                                     
\hline
A &  60565571641856 &  1.03$\pm$0.05 & -30$\pm$1 & -5.321$\pm$0.076 & -3.518$\pm$0.065 & 48.65$\pm$1.12 & -0.74$\pm$0.74 &-7.12$\pm$0.82 & 49.17$\pm$1.11 \\ 
B &  60569875463936 &  0.49$\pm$0.04 & -45$\pm$8 & -4.505$\pm$0.063 & -0.507$\pm$0.049 & 71.69$\pm$5.08 & -5.10$\pm$7.15 & 5.72$\pm$0.53 & 71.91$\pm$5.07 \\ 
C1&  60359417132800 &  5.31$\pm$0.48 & -40$\pm$6 & -2.415$\pm$0735  & -0.206$\pm$0.576 & 33.22$\pm$3.06 &-23.00$\pm$5.20 & 6.29$\pm$0.54 & 33.81$\pm$3.05 \\
D &  60363709280768 &  1.62$\pm$0.32 & -58$\pm$ 0.8&-4.566$\pm$0.636 & -2.248$\pm$0.376 & 52.16$\pm$2.73 &-30.83$\pm$1.61 & 0.64$\pm$1.68 & 52.16$\pm$2.73 \\
E &  60565573859584 &  0.14$\pm$0.22& -33$\pm$18 & 0.232$\pm$0.377 & -0.699$\pm$0.265 & 22.80$\pm$17.45 &-18.87$\pm$17.43 &-18.20$\pm$42.45 & 29.18$\pm$28.82 \\
F &  60569875460096 &  0.47$\pm$0.08 & -41$\pm$11 &-5.739$\pm$0.130 & -0.292$\pm$0.097 & 82.40$\pm$10.16 & 5.63$\pm$10.72  & 9.19$\pm$1.02 & 82.91$\pm$10.10 \\
G &  60359413315328 &  0.51$\pm$0.12 & -87$\pm$0.5&-4.417$\pm$0.191 & -4.064$\pm$0.143 & 93.01$\pm$8.85 & -40.33$\pm$5.37 &-28.20$\pm$8.28 & 97.19$\pm$8.80 \\
H &  60599931508480 &  0.62$\pm$0.20 & -78$\pm$10& -4.839$\pm$0.341 & -0.577$\pm$0.248 & 82.61$\pm$11.38& -36.89$\pm$10.49& 4.67$\pm$2.02 & 82.74$\pm$11.36 \\
J &  60565571749760 &  0.13$\pm$0.24& -52$\pm$6 & -3.900$\pm$0.373 & -1.054$\pm$0.292 &159.03$\pm$214.99& 38.04$\pm$125.88 &-17.10$\pm$45.78 &159.94$\pm$213.81 \\
N &  60565571767552 &  0.25$\pm$0.10 & -37$\pm$6 &  0.092$\pm$0.148 &  0.134$\pm$0.121 & 28.12$\pm$4.02 & -21.18$\pm$5.41 & 8.71$\pm$2.25 & 29.44$\pm$3.90 \\
O &  60569875457792 &  1.17$\pm$0.05 & -22$\pm$7 &  2.607$\pm$0.098 &  2.108$\pm$0.076 & 12.57$\pm$3.57 & -13.00$\pm$6.07 & 14.12$\pm$0.47 & 18.90$\pm$2.40 \\
P1&  60565571767424 &  0.17$\pm$0.09 & -43$\pm$10& -0.889$\pm$0.139 & -0.389$\pm$0.106 & 55.13$\pm$13.04& -11.71$\pm$11.20& -2.17$\pm$6.32 & 55.17$\pm$13.03 \\
U &  59092406721280 &  0.50$\pm$0.07 & -45$\pm$4 & -1.877$\pm$0.113 & -5.096$\pm$0.083 & 52.70$\pm$3.32 & -14.82$\pm$3.86 &-39.77$\pm$6.62 & 66.02$\pm$4.78 \\                  
\hline  
\end{tabular}
\end{center}
\end{table*}

\section{Star's orbits}

\noindent
Using the {\it Gaia} DR2 data system, we can calculate 
the orbits of the stars in the Galaxy. 
With known distances, proper motions can be translated into tangential 
velocities. From that and from the radial velocities already obtained, 
the total velocities of the targeted stars are reconstructed and their
orbits as they move across the Galaxy can then be calculated. 

\noindent
The stellar orbits are obtained by integration of the equations of 
motion. A 3D potential of the Galaxy is required for that. Here we 
use an axysimmetric potential
consisting of a spherical central bulge, a disk and a massive 
spherical halo, developed by Allen \& Santill\'an (1991). It is an 
analytic potential that allows an efficient and accurate numerical
computation of the orbits. In the present case, the total mass of
the Galaxy is assumed to be 9$\times10^{11}$ M$_{\odot}$.  We take the Sun
to be at 8.5 kpc from the Galactic center and moving circularly
with a frequency $\omega_{\odot}$ = 25.88 km s$^{-1}$ kpc$^{-1}$.
We do not consider a Galactic bar nor a spiral arms potential.

\begin{figure*}
\centering
\includegraphics[width=0.90\columnwidth]{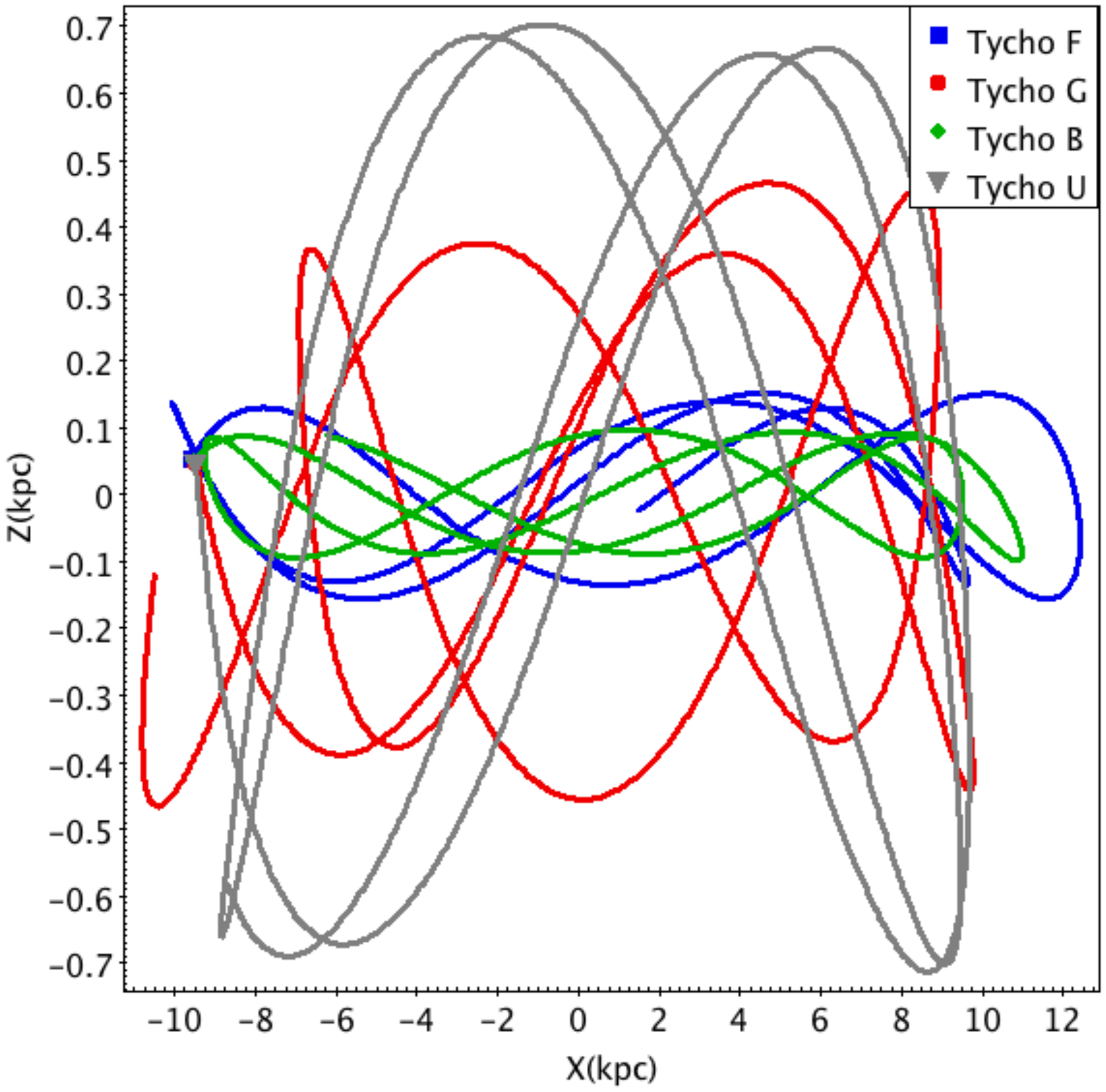}
\includegraphics[width=0.90\columnwidth]{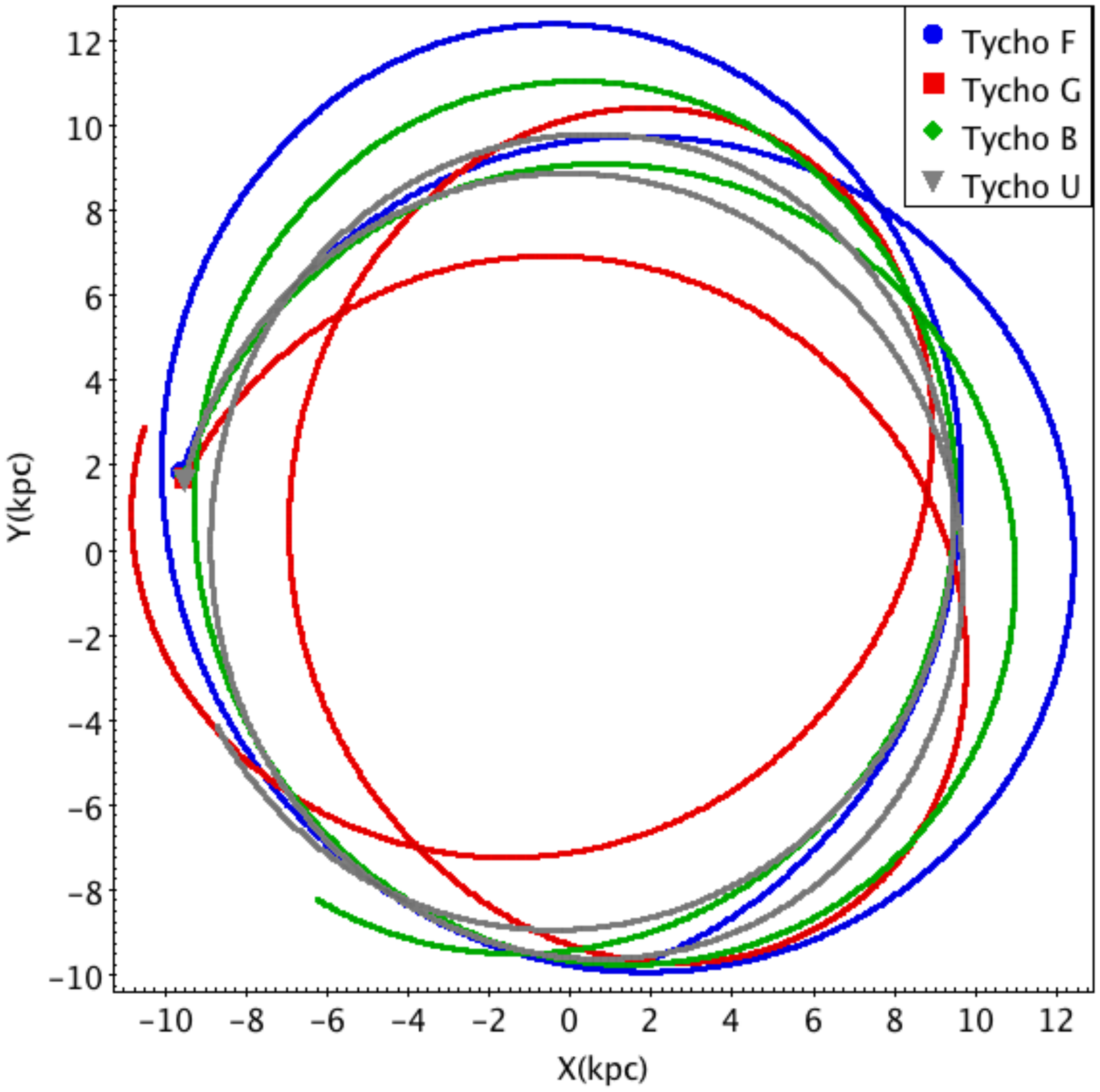}
\caption{The orbits of stars B (green), G (red), F ((blue), and
U (gray), projected on the Galactic meridian plane (left) and on the Galactic 
plane (right), computed forward on time for the next 500 Myr. 
The common starting point is marked with 
a blue square. In the left panel we  see that star U reaches the largest 
distance from the Galactic plane, followed by star G (which corresponds to the 
respective values of the W component of their velocities in Table 4), while 
stars B and F scarcely depart from the plane. The behaviour of the latter 
stars is typical of the rest of the sample considered here. In the right panel 
we see that the orbit of star G, on the Galactic plane, is highly eccentrical 
(which corresponds to the high value of the U component of its velocity in 
Table 4), while the other stars (including star U) have orbits close to 
circular. Also here, the behaviour of stars B and F is representative of the 
whole sample.}
\label{Figure 4}
\end{figure*}

\noindent
In Figure 4 we show the orbits of stars B, G, F and U. We see that only
stars G and U do reach large distances above and below the Galactic 
plane, while the other two stars, in contrast, do not appreciable
leave it. It can also be noted, in the motion parallel to the Galactic 
plane, the large eccentricity of the orbit of star G.

\noindent
The Figure is meant to show how far from the Galactic plane do reach the stars 
with significant proper motions in $\mu_{\alpha}$ and $\mu_{\delta}$. 
We take four stars with distance compatible with that of the SN. 
We see that star G will reach up to 500 pc and star U up to almost 700 pc above 
the Galactic plane within the next 500 Myr. In contrast, Tycho B and Tycho F 
(which have an insignificant $\mu_{\delta}$) do not reach 200 pc in their orbits
 in 
any lapse of time. We have shown Tycho B and Tycho F, because those have 
$\mu_{\alpha} \sim$ 4 mas yr$^{-1}$ but negligible $\mu_{\delta}$. Their orbits 
do not look peculiar. These are example of the many stars in a similar 
situation, which can be seen in our Tables. They will be thin disk stars (as 
seen in the Toomre diagram). The case of star U is unique, in the sense that 
it has a slighly larger $\mu_{\delta}$ than Tycho G. Tycho U has a negligible  
$\mu_{\alpha}$. This makes its orbit very circular. Tycho G has about the same 
proper motions in $\mu_{\alpha}$ and in $\mu_{\delta}$. This is why it 
reaches 500 pc above the Galactic plane but, at the same time, unlike star U, 
its orbit is eccentric. 

\noindent
The total velocity of star G is larger than that of star 
U. This can already be seen from the orbit and more explicitly in the Toomre 
diagram.  When we add the radial velocity vector to obtain the total velocity 
for star G, we have a v$_{r}$ = -87.40 km s$^{-1}$ (heliocentric), which is 
larger than that of star U (-45.40 km s$^{-1}$). Thus total velocity for star G 
is 103.69 km s$^{-1}$ while for U is only 68.63 km s$^{-1}$. 

\begin{figure*}
\centering
\includegraphics[width=0.90\columnwidth]{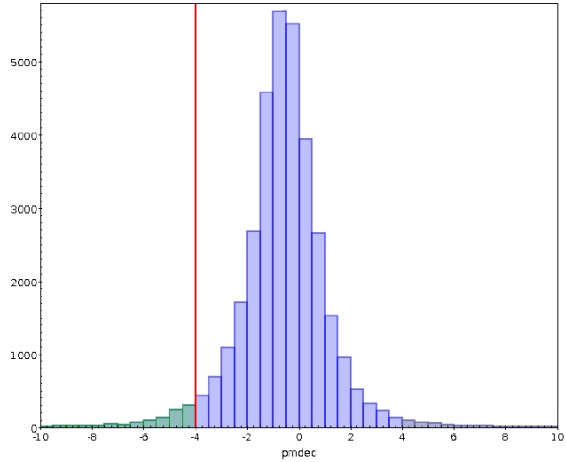}
\caption{ 
Histogram of the distribution in $\mu_{\delta}$ (in mas yr$^{-1}$) of 
the stars within 1 degree of the geometrical center of Tycho's SNR in the 
range of distance compatible with SN 1572 (1.7 $<$ d $<$  3.7 kpc). The data
 are 
obtained from  
{\it Gaia} DR2. The red vertical line shows the  $\mu_{\delta}$
 of star G.}
\label{Figure 5}
\end{figure*}

\begin{figure*}
\centering
\includegraphics[width=0.9\columnwidth]{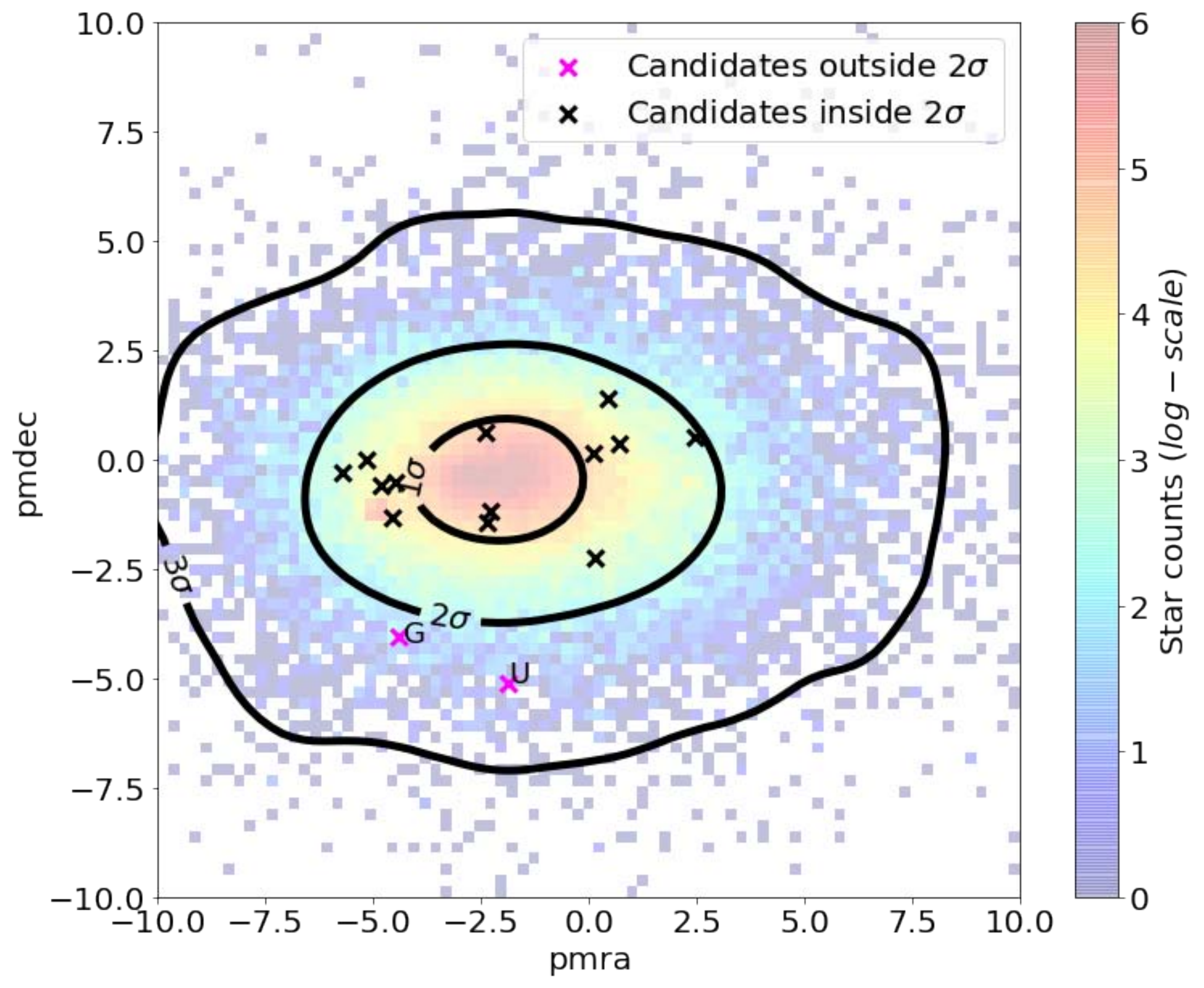}
\caption{
Proper motions of the candidate stars (see first column of Table 5) 
to companion 
  of SN 1572 ploted over the distribution of all the stars with 
distance compatible with that  of the SNR (1.7 $<$ d $<$ 3.7 kpc) and 
included 
within a radius of 1 degree of
the remnant.}
\label{Figure 6}
\end{figure*}

\noindent
In Figure 5, we have made an histogram of the heliocentric radial 
velocities of the stars at distances between 1.7 $<$ d $<$ 3.7 
within 1 degree from the geometrical center of the SN. 
We see there that the position of star G is anomalous. We also see that around 
1 degree in the SN position the heliocentric radial velocities are small. 

\noindent
 In Figure 6 we show the position of the stars compatible 
with the distance of SN 1572 in  proper motions in right ascension
and declination. 
All stars are those from
 the Gaia DR2  within 1 degree from the SN 1572 center and
 1.7 $<$ d $<$ 3.7 kpc.
 It is 
worth noting the logarithmic scale. Basically within 1 degree of the center
of SN 1572 and those distances
 the large majority of stars have very low proper motions.

\noindent
Readers  might ask what happens with the stars, in the full sample of Tables 2 
and 3, that do not have measured heliocentric radial velocities because they 
were far from the 15\% of the radius of the remnant explored in RL04. The {\it 
Gaia} DR2 data show no significant proper motions for any of them.

\section{ Candidate stars}

\noindent
In order to evaluate the likelyhood  that a given star were the companion of 
the SN, we look at the distances provided by {\it Gaia} parallaxes  and to
the proper motions. For some of the stars, we also have radial velocities,  
obtained from high resolution spectra. Parallaxes and proper motions allow 
to discern whether a given star has received extra momentum from the disruption
of a binary system to which it belonged. 
Such extra momentum (relative to the motion of the center of mass of the
binary) mostly comes from the orbital motion of the star when the system
was disrupted by the SN explosion. There is also some kick due to the
impact of the SN ejecta on the companion, but it is comparatively small
and depends on orbital separation and on how compact the companion is.
In the hydrodynamic simulations of Marietta, Burrows \& Fryxell (2000), the
momentum gained from the kick ranges from 12\% to 50\% of the momentum the star
had before explosion, for main-sequence and subgiant companions, the kick
being much smaller in the case of red giants. Similar values are found by
Pan, Ricker \& Taam (2014).

\noindent
We examine first the stars closest to the geometrical
center of the remnant. 

\bigskip
\noindent{\bf Stars A to G}

\noindent
Star A is the closest star to the geometrical center of the SN remnant. 
From its stellar atmosphere parameters, this star is at the foreground of
the SN. The {\it Gaia} parallax places it at 0.97 $^{+0.05}_{-0.04}$ kpc. 
By comparison of the absolute magnitude corresponding to the stellar 
parameters with the photometric data, we had derived $d$ = 1.1$\pm$ 0.3 kpc 
(R04, B14).

\noindent
 Star B has recently been thought to be a  foreground star 
(Kerzendorf et al. 2018). It is a hot star close 
to the geometrical center of the remnant.
{\it Gaia} DR2 places this star at a distance compatible with that
of the SNR.
$d$ = 2.03$^{+0.19}_{-0.15}$ kpc, is now compatible with the SN distance. 
We already pointed out, in R04 and B14, that the star likely was at a 
distance compatible with that of the SNR.
 We can discard star B, however, on the basis of having no peculiar 
proper motions nor radial velocity. We have made a reconstruction of the 
orbit of star B and it moves on the Galactic plane without any disturbance 
towards upper or lower Galactic latitudes.

\noindent
Stars C1, C2 and C3 have been observed by the {\it HST}. {\it Gaia} could 
only observe C1, and  has determined it to be a very nearby star, at a 
distance of 0.18$^{+0.03}_{-0.01}$ kpc. From the stellar parameters,  
a distance of 0.75$\pm$0.5 kpc was estimated. For C2 and C3,  RL04 and B14
 could not estimate 
any distance, due to their faint magnitudes. The proper motion values  
for these three stars obtained from {\it HST}  images in B14 show that they are  moving close to the Galactic plane. 

\noindent
Star D is also very nearby. It is at a distance of 0.62$^{+0.15}_{-0.11}$ kpc
according to {\it Gaia}. We had calculated a distance of $d$ = 0.8$\pm$0.2 kpc. 
Most of the targeted stars close to the geometrical center of the SNR 
are at distances below 1.5 kpc. 

\noindent
Star E, though, is at a very large distance. {\it Gaia} indicates a $d$ = 
7.22$^{+xxx}_{-4.43}$ kpc (the upper limit correponds to a negative parallax). 
Star E was suggested as the SN companion by Ihara et al. (2007).
 The authors detected 
absorption lines in the blue side, of the spectrum, at 3720 \AA, consistent 
with Fe absorption from Tycho's SNR. They concluded that this might either 
be due to the Fe I in the SN ejecta or to a peculiarity of the star. 
In GH09 it was pointed out that it is likely a very distant star.
In fact, both its distance and its kinematics do exclude it as a 
possible companion of the SN.

\noindent
Star F is compatible with the distance to Tycho's SN. {\it Gaia} measures 
a distance of $d$ = 2.15$^{+0.44}_{-0.32}$ kpc. We had calculated a distance 
$d$ = 1.5$\pm$0.5 kpc. It is not moving at high radial velocity nor
does it have a high proper motion perpendicular to the Galactic plane. 
Its orbit does not depart from the Galactic plane.

\noindent
Star G can also be considered among those close the center of the
SNR. {\it Gaia} has measured a distance $d$ = 1.95$^{+0.6}_{-0.35}$ kpc, 
which is within the range of distances suggested for the Tycho SN.  This
star has been a proposed companion in R04, GH09, and B14. Its 
kinematics corresponds to that of  a thick disk star, 
but it 
has a thin disk composition. It was found an  enhanced $[Ni/Fe]$ in GH09 and 
questioned in K13. A new calculation was done in B14, which still shows a 
value $[Ni/Fe]$ larger than the solar. We leave this point aside and refer
only to the agreed solar metallicity of the star.

\noindent
All the previous stars have measured v$_{r}$. They have been placed in the
Toomre diagram (Figure 3 and Table 4). 

\bigskip
\noindent{\bf Proposed stars at the NW of the geometrical center}

\noindent
Xue and Schaefer (2015) place the site of the explosion of Tycho's SN at 
the NW of the geometrical center of the SNR. They base their claim, in
part, on  a reconstruction of the historical center using observations of 
astronomers that wrote records on SN 1572 in the year of its discovery. 
Their position is at odds with a previous historically based 
reconstruction of the location in the sky of SN 1572 from Stephenson \& Clark 
(1977). On the other hand, Xue and Schaefer (2015)
 use a substitute of a 2D 
hydrodynamical simulation that, as noted by Williams et al. (2016),  
would only be valid for perfectly spherical remnants.

\noindent
Xue \& Schaefer (2015) 
give as the position of the explosion site  R.A = 00$^{h}$ 
25$^{m}$ 15.36$^{s}$ and Dec= 64$^{o}$ 08' 40.2''. 
From it, they suggest  that the companion star should be in a small circle 
around stars O and R. In their Figure 2 they point to stars  O, Q, S and R. 
From the {\it Gaia} DR2 data, we see that stars O, Q and S are at too short 
distances, incompatible with the distance to Tycho's SN.  Star O is at a 
distance of 0.85$^{+0.54}_{-0.22}$ kpc. Star Q is at a distance of 
1.51$^{+1.53}_{-0.51}$ kpc, only the  upper limit being compatible with the 
distance to the SNR, but it has no high proper motions. Star S is also at a 
small distance of 0.81 $^{+0.41}_{-0.20}$ kpc. Star R has no distance 
measurement nor proper motions in the {\it Gaia} DR2. However, we have 
proper motions measured with the {\it HST} (B14) and they are 
small.
 Star P is, as seen in the images of the {\it HST},
 two stars P1 and P2. In 
the {\it Gaia} DR2 we have only the P1 parallax, though not very well determined
but no proper motions are given. In B14, though, we have the proper motions 
of the stars and they are relatively small. P1 has a v$_{r}$ of --43 $\pm$ 10
and it has been possible to place it in the Toomre diagram (Figure 3 and 
Table 4) where it shows no kinematic peculiarity.  Star N is in the same 
 area of he sky as stars O, P1 and P2, Q, R and S. It 
has small proper motions and it is possible to place it in the Toomre diagram,
 where it lies in the region of low kinematical values. 
 In that corner of the sky suggested by Xue \& Schaefer (2015), 
there is no star looking as a companion of Tycho's SN in any way.

\bigskip
\noindent{\bf The NE proposed center}

\noindent
In a recent paper by Williams et al. (2016), the expansion center of
the remnant is suggested to be at the NE of the geometrical center. These 
authors use an extrapolation of the trajectories of different regions 
of the SNR,  but also a 
2D hydrodynamical simulation of the expansion of the ejecta in an inhomegeneous 
medium. They also assume cylindrical symmetry in the initial ejection of the 
supernova material. However, Krause et al (2008), from the spectrum of the 
light echo of SN 1572, suggest that the explosion was aspherical and thus 
not cylindrically symmetric. 

\noindent
Their suggested center is at R.A. = 00$^{h}$ 25$^{m}$ 22.6$^{s}$ and 
Dec = 64$^{0}$ 08$^{'}$ 32.7$^{''}$. This would be close to stars L and K. 
L is a star at a distance of $d$ = 1.45$^{+2.87}_{-0.58}$ kpc, but  with
small proper motions. K has no distance determined by {\it Gaia}. 
In B14 we suggest it to be around 4 Kpc. The kinematics  of the star, with 
small proper motions, makes it a non--suitable companion of Tycho's SN.  
In the NE center there are other stars like W, AK, AL, and AM. These stars
 do not have accurate parallaxes in the {\it Gaia} DR2 release. They
 have upper limits
in distance with negative parallaxes. We have looked  at their kinematics 
and they have moderate and low proper motions.

\noindent
Therefore, given the various candidates proposed, the best approach is to 
look for those that are within the range of the possible distance to 
Tycho's SN, show a peculiar kinematics and are within the region of the 
sky already explored.  

\noindent
We have placed in Table 5 the different stars and evaluated their 
viability as possible companions.

\smallskip

\begin{table*}
\scriptsize
\begin{center}
\caption{Criteria satisfied by the stars in Table 1. $v$ refers to the
total velocity, $v_{t}$ to the tangential velocity, and $v_{b}$ to the 
velocity perpendicular to the Galactic plane}
\begin{tabular}{llll}

\hline
\hline
Star & 1.7 kpc $\leq d \leq$ 3.7 kpc  & High Velocity & High $v_{b}$ \\ 
 (1) &              (2)               &     (3)       &     (4)      \\          
\hline

A   & No ($d$ = 0.97$^{+0.05}_{-0.04}$ kpc)  & --- &
      ---  \\
B   & Yes ($d$ = 2.03$^{+0.19}_{-0.15}$ kpc) & No ($v$ = 72$\pm$5 km/s) &
      No ($v_{b}$ = -0.5 $\pm$ 0.6 km/s)  \\
C1  & No ($d$ = 0.18$^{+0.03}_{-0.01}$ kpc)  & --- &
      ---  \\
D   & No ($d$ = 0.62$^{+0.15}_{-0.11}$ kpc)  & --- &
      ---  \\
E   & No ($d$ = 7.22$^{+xx}_{-4.43}$ kpc) & ---  &
      --- \\
F   & Yes ($d$ = 2.15$^{+0.44}_{-0.32}$ kpc) & Yes ($v$ = 83$\pm$10 km/s)&
      No ($v_{b}$ = 3 $\pm$ 1 km/s)   \\
G   & Yes ($d$ = 1.95$^{+0.60}_{-0.35}$ kpc) & Yes ($v$ = 103$\pm$7 km/s)&
      Yes ($v_{b}$ = -33 $\pm$ 9 km/s) \\
H   & Yes ($d$ = 1.61$^{+0.79}_{-0.40}$ kpc) & Yes ($v$ = 91$\pm$11 km/s)&
      No ($v_{b}$ = -0.5 $\pm$ 2.5 km/s)  \\
I   & ---                                  & ---                       &
      ---                             \\
J   & No ($d$ = 7.46$^{+xx}_{-4.77}$ kpc) & --- &
      --- \\
K   & ---                                  & ---                      &
      ---                            \\
L   & Yes ($d$ = 1.45$^{+2.87}_{-0.58}$ kpc) & No ($v_{t}$ = 10$^{+20}_{-6}$ 
km/s)&  No ($v_{b}$ = -2 $\pm$ 8 km/s)\\
M   & ---                                  & ---                      &
      ---                           \\
N   & Yes ($d$ = 4.06$^{+2.57}_{-1.14}$ kpc) & No ($v$ = 36$\pm$4 km/s) &
      No ($v_{b}$ = 2 $\pm$ 3 km/s) \\
O   & No ($d$ = 0.85$^{+0.54}_{-0.22}$ kpc)  & --- &
      --- \\ 
P1  & No ($d$ = 5.96$^{+7.23}_{-2.11}$ kpc)  & --- &
      ---  \\
Q   & Yes ($d$ = 1.51$^{+1.53}_{-0.51}$ kpc) & No ($v_{t}$ = 24$^{+24}_{-8}$ 
km/s) &  No ($v_{b}$ = 10 $\pm$ 10 km/s)\\
R   & ---                                  & ---                      &
      ---                            \\
S   & No ($d$ = 0.81$^{+0.41}_{-0.20}$ kpc)  & ---
  & --- \\
T   & Yes ($d$ = 1.77$^{+1.86}_{-0.60}$ kpc) & No ($v_{t}$ = 30$^{+32}_{-10}$ 
km/s) & No ($v_{b}$ = 4 $\pm$ 3 km/s)\\
U   & Yes ($d$ = 1.98$^{+0.32}_{-0.24}$ kpc) & No ($v$ = 68$\pm$5 km/s) &
      Yes ($v_{b}$ = -46 $\pm$ 8 km/s)\\
V   & No ($d$ = 16.81$^{+xx}_{-15.89}$ kpc)& --- 
    & --- \\
W   & No ($d$ = 5.17$^{+xx}_{-3.07}$ kpc) & --- 
    & --- \\
X   & No ($d$ = 5.20$^{+xx}_{-3.59}$ kpc) & --- 
    & --- \\
Y   & Yes ($d$ = 1.58$^{+0.87}_{-0.41}$ kpc) & No ($v_{t}$ = 18$^{+10}_{-5}$ 
km/s) & No ($v_{b}$ = -17 $\pm$ 9 km/s)\\ 
Z   & No ($d$ = 5.68$^{+2.72}_{-2.57}$ kpc) & --- 
    & --- \\
AA  & Yes ($d$ = 1.04$^{+1.00}_{-0.33}$ kpc) & No ($v_{t}$ = 4$^{+8}_{-7}$ km/s)&
      No ($v_{b}$ = -5 $\pm$ 6 km/s)\\
AB  & ---                                  & ---                        &
      ---   \\
AC  & Yes ($d$ = 2.04$^{+0.99}_{-0.50}$ kpc) & No ($v_{t}$ = 11$^{+6}_{-4}$ km/s)&
      No ($v_{b}$ = -12  $\pm$ 7 km/s)  \\
AE  & Yes ($d$ = 3.58$^{+5.84}_{-1.37}$ kpc) & No ($v_{t}$ = 16$^{+13}_{-5}$ km/s)
    & No $(v_{b}$ = 6$^{+20}_{-1}$ km/s) \\
AF  & No ($d$ = 0.76$^{+0.24}_{-0.15}$ kpc)  & --- 
    & --- \\
AG  & Yes ($d$ = 1.42$^{+1.69}_{-0.50}$ kpc) & No ($v_{t}$ = 10$^{+12}_{-5}$ 
km/s) & No ($v_{b}$ = -6 $\pm$ 6 km/s)\\
AH  & No ($d$ = 4.85$^{+3.55}_{-1.44}$ kpc) & --- 
    & --- \\
AI1/HP-1 & No ($d$ = 0.35$^{+0.04}_{-0.03}$ kpc) & ---  
    & --- \\
AI2 & ---                                      & ---                    &
      ---    \\
AJ  & No ($d$ = 5.35$^{+xx}_{-2.81}$ kpc) & --- 
    & --- \\
AK  & ---                                      & ---                    &
      ---    \\
AL  & No ($d$ = 2.61$^{+xx}_{-1.61}$ kpc) & --- 
    & --- \\
AM  & No ($d$ = 1.33$^{+xx}_{-0.70}$ kpc)  & --- 
    & --- \\
AN  & Yes ($d$ = 1.65$^{+0.49}_{-0.30}$ kpc) & No ($v_{t}$ = 24$^{+7}_{-4}$ 
km/s) & No ($v_{b}$ = -7 $\pm$ 3 km/s)\\

\hline

\end{tabular}
\end{center}
\end{table*}

\smallskip

\section{Luminosities and  models}

\noindent
There are significant differences in the predictions of the characteristics of 
the surviving companions of the supernova explosion. Podsiadlowski (2003) found 
that, for a subgiant companion, the object $\sim$ 400 years after the explosion 
might be either significantly overluminous or underluminous, relative to 
its pre-SN luminosity, depending on the amount of heating and the amount 
of mass stripped by the impact of the SN ejecta. More recently Shappee, 
Kochanek \& Stanek (2013) have also followed the evolution of luminosity
for years after the impact of the ejecta on a main--sequence the companion. The 
models first rise in temperature and luminosity, peaking at 
10$^{4}$ L$_{\odot}$ to start cooling and dimming down to 10 L$_{\odot}$ some 
10$^{4}$ yr after
 the explosion. Around 500 days after explosion the companion 
luminosity would be 10$^{3}$ L$_{\odot}$. Pan, Ricker \& Taam (2012, 2013, 
2014) criticize the two preceding approaches for the arbitrary of the initial 
models. Starting from their hydrodynamic 3D models, they find lower 
luminosities for the companions than the previous authors. They find 
luminosities of the order of only 10 L$_{\odot}$ for the companions, several 
hundred days after the explosion. 

\noindent
Now, knowing the distances from {\it Gaia}, we can derive the luminosities of
the stars compatible with being inside the SNR.  We take a distance
 to the SNR coming from the measurements from various reliable approaches,
 which means a value 1.7 $<$ d $ <$ 3.7 kpc. We have 15 candidates 
compatible with that distance. We already had {\it UBV} photometry for some of
them and now {\it Gaia} photometry for all. From that we find that there
is no clearly overluminous candidate. 

\noindent
It has been suggested that, within the double--degenerate channel to produce 
SNe Ia, the explosion can be triggered just at the beginning of the coalescence
process of the two WDs, by detonation of a thin helium layer coming from the 
surface of the less masive one. That would induce a second detonation in the
core of the more massive WD. This hypotetical process has been dubbed as the 
``dynamically driven double--degenerate double--detonation scenario'' (see 
Shen et al. 2018 and references therein). In this case, the less massive 
WD would survive the SN explosion and be ejected at the very high orbital 
velocity ($>$ 1000 km s$^{-1}$) it had at the moment of the explosion. Those
would be seen as ``hypervelocity WDs'' (Shen et al. 2018). The number of 
hypervelocity WDs detectable by {\it Gaia} depends on the assumed luminosity 
of these objects. Shen et al. (2018) conclude that, taking into account tidal 
heating undergone by the WD before the explosion, a typical object would have, 
after subsequently cooling for $\sim$ 10$^{6}$ yr, a luminosity $\geq$ 0.1 
$L_{\odot}$, and thus be detectable by {\it Gaia} up to a distance of 1 kpc. 
Based on that, they predict that $\sim$ 30 potentially detectable hypervelocity 
WDs should be found within 1 kpc from the Sun. They have actually found, 
from {\it Gaia} DR2, three objects that, after having been followed up with 
ground--based telescopes, although not looking as typical WDs might be the 
result of heating and bloating of a SN Ia WD companion.       

\noindent
In the case of Tycho's SN, 
the cooling time of a possible surviving WD companion is only $\sim$ 450 yr, 
and thus the luminosity should be significantly higher than the  
0.1 $L_{\odot}$ adopted by Shen et al. (2018) for a typical companion having
cooled for $\sim$ 10$^{6}$ yr. 

\noindent
In order to look for a possible hypervelocity WD companion to Tycho, we 
must considerably enlarge the search area around the center of the SNR. 
Taking as an upper limit a velocity perpendicular to the line of sight of 
4000 km s$^{-1}$, the maximum distance traveled in 450 yr, 5.7$\times 10^{13}$
km, translates, at a distance of the SNR, into an angular 
displacement of 2.1 arcmin (that is slightly more than 50\% of the average 
radius of the SNR, which is about 4 arcmin).

\noindent
We have checked that there is no object with unusually high proper motion
in the {\it Gaia} DR2 data release, within the searched area and up to a 
$G$--magnitude of 20.7 (V $\sim$ 22).
 For an extinction $A_{V}$ = 2.4 mag (GH09), and at 
the distance of Tycho, that means a luminosity $L \sim 0.3 L_{\odot}$, 
similar to the lower limit adopted by Shen et al. (2018). That does not take 
into account the capture of radioactive material by the companion WD predicted 
by Shen \& Schwab (2017). Objects such as the three candidates to hypervelocity 
former SN Ia companions found by Shen et al. (2018), with G--magnitudes 
$\sim$ 17--18 mag, would be clearly seen.

\section{Summary and conclusions}

\noindent
We have reexamined the distances and proper motions of the stars close
to the center of Tycho's SNR, using the data provided by the {\it Gaia} 
DR2. Previously, the distances were only know from determination of 
the stellar atmosphere parameters and comparison of the corresponding 
luminosities with the observed apparent magnitudes, with only an 
approximative knowledge of the extinction and uncertainty about the
luminosity classes in a number of cases. More accurate were the proper
motions, coming from astrometry made with the {\it HST}, but the 
DR2 has allowed a cross--check here. Besides, only a precise knowledge
of the distances allows to convert proper motions into tangential velocities
reliably. 

\noindent
{\it Gaia} now provides the last word about the distances and kinematics 
of the previously proposed companions of Tycho's SN. 

\noindent
A good agreement between the distances from {\it Gaia} DR2 and those
reported in B14 has been found in many cases, but with
a general trend to shorter {\it Gaia} distances as compared with
B14, which can be attributed to an underestimate of extinction in 
the direction of the remnant, in B14. In a few cases, however, 
the discrepancies are large.

\noindent
Concerning proper motions, the agreement is very good once due
account is made of the systematic effect of the motion of the 
local frame to which the {\it HST} measurements are referred 
with respect to {\it Gaia}'s absolute frame.  

\noindent
We find that, within the remaining uncertainties, up to 15
stars are at distances compatible with that of the SNR. The case
for Tycho G is that in samples such as the one shown in Figure 3, 
this star has a thick disk kinematics, but has thin 
disk metallicity. There is only a 0.8\% of star having similar characteristics.
We have inspected the proper motions of all the stars visible up to limit of 
the Gaia DR2 in magnitude,
and we have found no one with the same peculiar total velocity. 
We have presented a scenario in which Tycho G could  be the companion of 
the SN 1572. 
There is, 
however, the possibility that after performing several orbits around the 
Galactic center, and encountering globular clusters and spiral arms, the 
star orbit becomes eccentric and migrates towards higher Galactic latitudes. 
This is a suggested explanation for the characteristics of Tycho G.
A counterargument is why the other stars  that could have performed as well
 several Galactic orbits, in close locations, would not have 
migrated. 

\noindent
We agree
with Kerzendorf et al. (2018) that Tycho B is not a good candidate to
companion of the explosion. We can also exclude, in view of the {\it Gaia} DR2
data, that star E could be a companion, since it lies very far away. 

\noindent
In case that Tycho G were not the companion star, the double--degenerate 
scenario or the core degenerate scenario are favored, since we have gone well 
below solar luminosities.

\noindent
With {\it Gaia} DR2, we have also looked for the hypervelocities stars 
predicted by some scenarios, but within the magnitudes reached by {\it Gaia} 
we have found none.

\bigskip
\noindent
\section{Acknowledgements}

 This work has made use of data from the European Space Agency
(ESA) mission {\it Gaia} (https://www.cosmos.esa.int/gaia), processed by the
{\it Gaia} Data Processing and Analysis Consortium (DPAC, 
https//www.cosmos.esa.int/web/gaia/dpac/consortium). Funding for the DPAC has
been provided by national institutions, in particular the institutions 
participating in the {\it Gaia} Multilateral Agreement.  
P.R.--L. is supported by 
AYA2015--67854--P from the Ministry of Industry, Science and Innovation of 
Spain and the FEDER funds. J.I.G.H. acknowledges financial support from the
Spanish MINECO (Ministry of Economy of Spain) 
 under the 2013 Ram\'on y Cajal program MINECO RyC--2013--14875,
also from the MINECO AYA2014--56359--P and AYA2017-86389-P. 
This
work was supported as well by the MINECO through grant ESP2016--80079--C2--1--R 
(MINECO/FEDER, UE) and ESP2014--55996---C2--1--R (MINECO/FEDER, UE) and 
MDM--2014--0369 of ICCUB (Unidad de Excelencia 'Maria de Maeztu).

\end{document}